# Optical Torque from Enhanced Scattering by Multipolar Plasmonic Resonance

*Yoonkyung E. Lee[1], Kin Hung Fung[1,2], Dafei Jin[1], Nicholas X. Fang[1,*]*

[1]Department of Mechanical Engineering, Massachusetts Institute of Technology, Cambridge, MA 02139, USA

[2]Department of Applied Physics, The Hong Kong Polytechnic University, Hong Kong

nicfang@mit.edu

## ABSTRACT

We present a theoretical study of the optical angular momentum transfer from a circularly polarized plane wave to thin metal nanoparticles of different rotational symmetries. While absorption has been regarded as the predominant mechanism of torque generation on the nanoscale, we demonstrate numerically how the contribution from scattering can be enhanced by using multipolar plasmon resonance. The multipolar modes in non-circular particles can convert the angular momentum carried by the scattered field, thereby producing scattering-dominant optical torque, while a circularly symmetric particle cannot. Our results show that the optical torque induced by resonant scattering can contribute to 80% of the total optical torque in gold particles. This scattering-dominant torque generation is extremely mode-specific, and deserves to be distinguished from the absorption-dominant mechanism. Our findings might have applications in optical manipulation on the nanoscale as well as new designs in plasmonics and metamaterials.

# 1 INTRODUCTION

Optical force arises when the linear momentum carried by light is transferred to matter through both absorption and scattering. [1,2] When absorption dominates, each photon absorbed by the medium transfers a finite quanta of linear momentum $\hbar \mathbf{k}$, [3,4] where $\hbar$ is the reduced Planck's constant and $\mathbf{k}$ is the wavevector of the incoming light. When scattering dominates, on the other hand, the total transferred linear momentum from each photon is determined by the difference of the wavevectors before and after scattering $\mathbf{k} - \mathbf{k}'$. [5] The transfer of momentum through scattering is therefore not necessarily directly proportional to the number of photons scattered, which is the fundamental reason why it is possible to realize unconventional ways to transfer linear momentum from light to matter such as the recent example of an optical 'pulling' force. [6,7]

Investigations on optical manipulation [8,9] have been extended to the control of optical torque and optical angular momentum. [10,11] It has been well known that light carries both spin angular momentum (SAM) [12,13] and orbital angular momentum (OAM). [14,15] Both can be used to rotate small particles. While the optical torque from absorption is proportional to the number of photons absorbed, torque from scattering can be modified with greater freedom as in the case of linear force. The essential requirement to create torque predominantly from scattering is that the scattered light should be made to exhibit a twist in the azimuthal direction, meaning that the angular momentum carried by the scattered field should be different from that of the incident field. This has been widely explored using microscale dielectric particles that have negligible absorption under stable optical trapping conditions. [16] One way to produce the necessary twist is to shine a carefully aligned beam that carries OAM. [17] On the other hand, when the incident light carries only SAM or has no angular momentum to begin with, the twist needs to be provided by the optical response of the object; for example, the intrinsic [18–20] and mimicked [21] anisotropy of the material, as well as the asymmetry of the structure, such as chirality or structural windmill effects. [22–25] Very recently, it has also been theoretically predicted that light can exert "negative optical torque" on aggregates of dielectric microspheres. [26]

In contrast, the manipulation of subwavelength particles remains a more challenging issue: dielectric nanoparticles have negligibly small optical responses, and absorptive nanoparticles experience radiation pressure and thermal effects that often predominate the strength of conventional optical traps. [27,28] Despite such restraints, researchers have developed various ways to produce optical torque on the nanoscale: the principles in creating structurally asymmetric metallic rotors [29,30] have been scaled down to the nanoscale by enhancing the optical response of subwavelength particles through the excitation of surface plasmon (SP) resonance. [31,32] Researchers have also demonstrated KHz-rate spinning of gold nanospheres in water by the absorptive transfer of SAM from circularly polarized light. [33] SP resonance has also been widely used in designing nanoscale optical elements, inside systems for the spatial modulation of light such as optical nano-tweezers, [34–38] optical antennas, [39,40] and OAM-mediating metasurfaces. [41,42] In fact, the contributions from absorption and scattering are often left undistinguished in the discussion of SP-enhanced mechanical effects. To our knowledge, there have been no systematic studies so far on the optical torque that is particularly associated with the scattering of light under plasmonic resonance.

This article presents the study on how light scattering at plasmonic resonance can provide an additional channel for the control of optical torque, in addition to the commonly acknowledged mechanism of absorption. Our numerical calculations are based on simple non-chiral geometries, via exciting the first few orders of multipolar SP resonance. In the literature, the phenomena of multipolar SP resonance of metallic nanoparticles have been widely reported for various symmetry-breaking geometries, [43–50] including the flat triangular nanoprisms [51–53] used in this work. We show that the angular momentum carried by the scattered field depends greatly on the azimuthal charge distribution on the plasmonic surface, which is governed by the order of the multipolar SP resonance. This scattering-dominant mechanism of torque generation is distinctly different from the conventional means of absorption-dominant torque generation. The efficiency of angular momentum transfer is analyzed with respect to variations in size and rotational symmetry.

## 2 CONCEPTS AND METHODS

Broken rotational symmetry fundamentally provides the selection rule not only for the emergence of multipolar resonance in plasmonic particles, but also for the possible channels of optical angular momentum conversion through light-matter interaction. We propose how structures supporting multipolar resonance can be designed to produce torque by dictating the azimuthal diffraction order of the scattered field.

Suppose that light carrying an azimuthal angular momentum of $\mathbf{J} = \hbar m \hat{\mathbf{z}}$ is normally incident on an optical system. In the cylindrical ($\hat{\mathbf{r}}, \hat{\varphi}, \hat{\mathbf{z}}$) coordinates, the azimuthal field direction follows $e^{im\varphi}$, where $m$ denotes the total azimuthal order number. The optical system supports a discrete $N$-fold rotational symmetry in $\hat{\varphi}$, *i.e.*, the optical properties of the system are identical with respect to a rotation of an angular period of $2\pi/N$. After the process of scattering, the azimuthal distribution of the scattered field can change to $e^{im'\varphi}$, where $m'$ is the total output azimuthal order of the scattered field that can be expressed as [55,56]

$$m' = m + jN, \, (j = 0, \pm 1, \pm 2, \cdots), \tag{1}$$

in which *j* is an integer representing the angular diffraction order. The change in the azimuthal order number is a result of light diffraction, and therefore can be expressed by an angular grating rule. Eq.(1) can be interpreted as the angular counterpart of the linear grating equation $k' = k + jK, (j = 0, \pm 1, \pm 2, \cdots)$, where $k = |\mathbf{k}|$ represents the linear momentum of the photon in the direction of the wavevector, and $K$ is the added lattice momentum when the optical properties of the system are identical with respect to a translation by a linear period (pitch) of $2\pi / K$.

While all light-matter interactions follows Eq.(1) in principle, the angular grating effect is negligible when there is no mechanism to couple the light into nonzero diffraction order $j$. Most optical systems that are larger than the wavelength lack a mechanism to selectively enhance a particular output mode, and the dominant response becomes $m = m'$ with a conserved azimuthal order.

On the other hand, multipolar SP resonance is inherently successful at converting the azimuthal phase distribution of the plasmonic nearfield into the azimuthal order $m'$ that matches the order of the multipolar resonance. Using this fact, a plasmonic structure possessing discrete rotational symmetry would serve as an effective 'angular diffraction grating' that can selectively enhance the conversion into a certain output mode with $m'$ by matching the wavelength of the surface plasmon mode with the characteristic length scale of the angular grating.

A gold triangle depicted in Figure 1A is a simple example of an angular grating possessing 3-fold discrete rotational symmetry. We choose to investigate the results when the incident electric field is a RCP circularly polarized plane wave, described as $\mathbf{E}(r, \varphi, z) = E_0 [\mathbf{e}_r + i\mathbf{e}_\varphi] e^{im\varphi} e^{ikz}$ with $m = +1$ for right-handed circular polarization (RCP). Cylindrical coordinates is used to describe light carrying angular momentum in its most natural form, where $\mathbf{e}_r$, $\mathbf{e}_\varphi$, and $\mathbf{e}_z$ are the unit vectors along the $r$-, $\varphi$-, and $z$-directions, respectively, and $m$ represents the projected total angular momentum onto the $z$-axis. Left-handed circular polarization (LCP) can be written as $\mathbf{E}(r,\varphi,z) = E_0[\mathbf{e}_r - i\mathbf{e}_\varphi]e^{im\varphi}e^{ikz}$, with $m = -1$. Recall that any plane wave can be expressed as a weighted sum of two orthogonal circular polarizations with a phase offset of $\pi/2$. Throughout this paper, an RCP plane wave incidence is assumed without the loss of generality.

After interacting with the gold nanoparticle, the scattered electric field of total azimuthal order $m'$ can be expressed as $\mathbf{E}'(r,\varphi,z) = [E_r(r,z)\mathbf{e}_r + E_\varphi(r,z)\mathbf{e}_\varphi + E_z(r,z)\mathbf{e}_z]e^{im'\varphi}$. The resultant azimuthal order number $m'$ is dictated by Eq.(1), which is $m' = m + 3j$ for 3-fold symmetry, where $j$ is an integer. Using RCP incidence, the azimuthal order can be changed into values, including $m' = -2$, which represents a quadrupolar azimuthal mode with an opposite direction of rotation, henceforth denoted as the negative quadrupole mode. Eq.(1) governs the possible orders of multipolar SP resonance that can be excited by a normally incident plane wave, which shows the natural connection between multipolar SP resonance and the possible channels of optical angular momentum transfer. Due to this change in the azimuthal distribution of the scattered field, the triangular particle is expected to experience a torque from scattering, as well as absorption. In Figure 1A, the circular arrows represent the existence of torque contribution from not only absorption (red) but also from scattering (blue).

In contrast, a circular disk illustrated in Figure 1B bears continuous rotational symmetry. Mathematically, one may treat the disk to have $N \to \infty$, and so any non-zero diffraction order will require $m' \to \infty$, which is practically always decoupled from the incoming light. The circular disk fails to modify the azimuthal order of the scattered field. Therefore we predict that no scattering torque would arise for a circular disk, as illustrated in Figure 1B.

As is well-known, the mechanical effects delivered by photons onto an object can be characterized via inspecting the photonic momentum flux around the object. For a mechanically fixed object exposed to a steady incident flux, the time-averaged total torque can be obtained from an area integral of the time-averaged Maxwell Stress Tensor (MST) $\vec{\mathbf{T}}$ over an arbitrary closed surface $S$ surrounding the object:

$$\mathbf{M}_{tot} = \oint_S (\mathbf{r} \times \vec{\mathbf{T}}) \cdot d\mathbf{A} = \mathbf{M}_{abs} + \mathbf{M}_{sca} . \qquad (2)$$

In general, $\mathbf{M}_{tot}$ can be separated into the absorption part $\mathbf{M}_{abs}$, and the scattering part $\mathbf{M}_{sca}$, as discussed further below. The component of the MST $\vec{\mathbf{T}}$ in Eq. (2) can be expressed as:

$$T_{\alpha\beta} = E_\alpha D_\beta + B_\alpha H_\beta - \frac{1}{2}\delta_{\alpha\beta}(\mathbf{E} \cdot \mathbf{D} + \mathbf{B} \cdot \mathbf{H}), \qquad (3)$$

where $\mathbf{E}$, $\mathbf{D}$, $\mathbf{B}$ and $\mathbf{H}$ are the electric field, the electric displacement field, the magnetic flux density, and the magnetic field, respectively, all taken from the total field data. A pair of indices $\alpha$ and $\beta$ denotes a particular component of momentum flux that points along the $\alpha$-axis and crosses the surface normal to the $\beta$-axis. In our calculation, the surface of integration is chosen as a rectangular box enclosing the gold nanoparticle, conforming to the Cartesian meshing condition used by the FDTD solver. Thus $\alpha$ and $\beta$ are coordinate indices for $x$, $y$, or $z$.

The distinction between the torque contributions from absorption and scattering can be made, by acknowledging that the absorptive torque is proportional to the number of photons absorbed. For the part of the incoming photons that are eventually absorbed by the object, the corresponding optical torque can be calculated from [57] as

$$\mathbf{M}_{abs} = \frac{C_{abs} I_{inc}}{\omega} m\hat{\mathbf{z}} . \qquad (4)$$

Here, $C_{abs}(\lambda) = P_{abs} / I_{inc}$ is the absorption cross section, in which $P_{abs}$ is the power absorbed by the object and $I_{inc}$ is the incident intensity. The number of absorbed photons per unit time is $P_{abs}/\hbar\omega$.

In all of the calculations, the incident intensity $I_{inc}$ is normalized to be 1. Physically, absorption is caused by material dissipation and is embedded into the calculation through the imaginary part of the dielectric function $\text{Im}[\varepsilon(\lambda)]$.

Since the value of the total optical torque can be numerically obtained through Eq. (2), the scattering contribution is calculated to be $\mathbf{M}_{sca} = \mathbf{M}_{tot} - \mathbf{M}_{abs}$. It is worthwhile to emphasize again that scattering torque cannot be calculated from a simple expression like Eq. (4), because the torque from scattering is not simply proportional to $C_{sca}$. The scattering cross section does not capture how the angular momentum of the scattered photons redistributes after scattering. A naïve calculation can cause an over- or under-estimation of torque.

All of the numerical results presented in this letter is calculated using the finite-difference time-domain (FDTD) method. [58,59] Without loss of generality, the metal used in this paper is gold (with permittivity [60] given in S1), the dielectric medium is air (with refractive index $n=1$), and incident light is modeled as a broadband pulse that is analyzed as a collection of monochromatic incidence. Linear scattering is assumed throughout the paper, meaning that the angular frequency $\omega$ of the photon does not change in the process of scattering.

**3 RESULTS AND DISCUSSIONS**

The numerically obtained torque spectra are plotted in Figure 2. The results are compared between three representative geometries possessing different orders of rotational symmetry: $N =$ 3, 4, and $\infty$. The material and the incident illumination are identical for all three cases, following the setup of Figure 1. The characteristic lateral size and the thickness are identically set to be 400 nm and 40nm, respectively.

Figure 2A shows the torque spectrum for a triangular plate with 3-fold discrete rotational symmetry. A clear enhancement of scattering torque (blue curve) is observed at the wavelength of the negative quadrupole mode ($m' = -2$). The distribution of the charges on the metal surface is represented by the colored electric field profiles plotted 4nm above the surface. The scattered quadrupole field rotates in the opposite direction from the incident field, because the possible orders after conversion is dictated by Eq. (1). As a result, an unusually large, scattering-dominant mechanical torque is created.

Figure 2B shows the consistent response of a square plate with 4-fold discrete rotational symmetry, which shows that the scattering torque is distinctly enhanced at the wavelength of the negative hexapole mode ($m' = -3$). Absorptive torque (red curve), on the other hand, does not exhibit sharp peaks, and shows a remarkable similarity between the three distinct geometries.

In stark contrast, the scattering torque is zero for the case of Figure 2C for the case of a circular disk with continuous rotational symmetry. No multipolar conversion is observed for the case of $N = \infty$. Since the azimuthal distribution of the scattered field resembles that of the incidence, scattering fails to create any torque.

The scattering contribution of torque by objects possessing discrete rotational symmetry originates from the constructive interference of the diffracted light. For subwavelength dielectric structures, the angular momentum conversion and hence torque generation via the scattering mechanism is quite small. However, the situation is rather different in metallic structures, where the conversion efficiency can be selectively enhanced by the corresponding order of multipolar SP resonance.

Traditionally, absorption has been regarded as the predominant mechanism of optical torque generation on plasmonic particles, especially when the transfer of SAM is concerned. Our results show that this is not always true. Scattering can become the predominant mechanism of torque generation, provided that the structure supports a conversion of azimuthal order between the incident and the scattered fields. Although in principle the scattered light with a converted azimuthal angular momentum cannot propagate to the far field, such a conversion is still capable of applying a significant torque onto the object in the near field.

To quantify the relative strength of scattering torque in the process, we define a dimensionless conversion ratio $\eta$ as

$$\eta = \frac{G_{sca}/\hbar m}{P_{ext}/\hbar \omega}. \tag{5}$$

In the numerator, $G_{sca}$ is the total angular momentum per unit time carried away by the scattered field, $\hbar m$ is the angular momentum of each incoming photon. In the denominator, $P_{ext}$ is the extinction power and $\hbar \omega$ is the energy of each extinguished photon. $\eta$ can be either positive or negative. (See S2 for the plot of extinction.)

In Figure 3, we show the calculated spectra of $\eta$ for the three particles analyzed in Figure 2. (black curves) The colored curves represent the effect of particle size. According to Eq.(5), $\eta = 1$ corresponds to the absence of conversion. This is observed for all of the curves in the vicinity of 900nm, which falls into the dipole resonance regime. On the other hand, $\eta = m'$ corresponds to the maximum conversion, when no absorption happens and all of the photons are converted from $m = 1$ into $m'$.

In contrast, the conversion ratios for the three geometries follow a clearly different trend in the 400~800nm wavelength range. In Figure 3A, the black curve corresponds to the dimensions of the particles in value of $\eta$ dips down below zero when the negative quadrupole mode is excited. This indicates that the scattered field may carry an angular momentum in the opposite direction to that of the incident angular momentum, i.e. $\eta < 0$. Under this circumstance, the object can experience an extraordinarily large torque. Figure 3B shows a similar trend, and the dips correspond to the negative hexapole mode of the square particle. The magnitude of the dip of the square particle is generally much smaller than that of the triangular particle, indicating a lower efficiency of angular momentum conversion. This is directly related to the quality of the SP resonance, which falls as the multipolar order increases. The size of torque depends not only on the efficiency of conversion but also on the size of the scatterer. In fact, the amount of torque is very sensitive to particle size, which explains why the size of torque is larger for the square in Figure 2, although the conversion is less efficient compared to the triangle.

For the circular disk in Figure 3C, $\eta$ does not contain any feature of resonant angular momentum conversion. The curve shows the baseline that represents the angular momentum altered by the absorption of gold. This absorptive baseline is very similar for all of the graphs in Figure 3, regardless of their size or rotational symmetry. Since $\eta$ quantifies the efficiency of angular momentum conversion between the incident and the scattered fields, the similarity in the absorptive baseline shows the limit for absorptive torque enhancement.

In each graph, the colored curves show the optimal size for reaching the best conversion efficiency. The optimum size can be estimated by comparing the incident wavelength with the circumferential length of the particle ($d \sim \lambda$). While the qualitative behavior is insensitive to small size variations, the relative size of the dips represents the difference in the quality of the multipolar SP resonance. When the particle size is made much smaller or much larger than the current settings ($d \sim \lambda$), the metallic particle can no longer support the high-quality multipolar SP modes, and the torque from scattering will be lost.

The importance of broken rotational symmetry can be explicitly appreciated in Figure 4. We show how $\eta$ varies when continuous symmetry is gradually broken, from 6-fold symmetry to a 3-fold symmetry. The change in the conversion ratio dip is indeed very sensitive to the symmetry of the geometry. In this series of geometry being studied, the lower the order of discrete rotational symmetry is, the more remarkable the resonant torque enhancement can be.

## 4 CONCLUSION AND OUTLOOK

Multipolar plasmonic resonance on symmetry-breaking nanostructures offers a new channel to boost the transfer of angular momentum, in addition to the previously acknowledged mechanisms of absorption. This paper discusses the concept of angular momentum transfer between photons and a flat nanostructure. The choice of cylindrical coordinates provides an elegant two-dimensional description of angular momentum transfer, since the projected angular momentum index $m'$ is itself sufficient to describe the in-plane torque on the flat particles with discrete rotational symmetry. Our next step is to generalize this approach to three-dimensional geometries, which involves field expansion in spherical coordinates. By doing so, we see a natural connection to the formalisms developed in quantum mechanics, where addition of angular momentum is commonly pursued using the formalism of Clebsch–Gordan coefficients.

The numerical simulation presented in this letter suggests that the SP-enhanced scattering can lead to a negative angular momentum conversion ratio, and hence produce an extraordinarily large torque. Although our analysis, which relies on the surface integration of the Maxwell's stress tensor, assumes that the surrounding fluid does not absorb light, our results could be a reasonable approximation to some experimental conditions. Our finding may lead to useful applications in the field of light-mediated mechanical manipulation as well as spatial light manipulation using metamaterial elements.

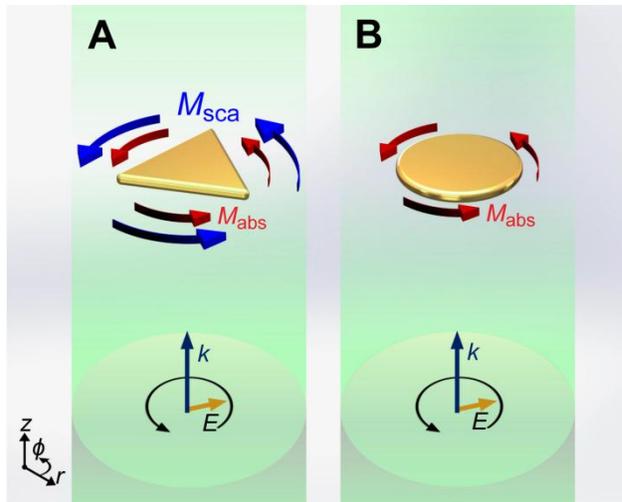

**Figure 1.** Two different mechanisms of optical torque generation mediated by absorption and scattering. A circularly polarized plane wave, denoted by the wavevector **k** and the electric field **E**, is normally incident on two gold nanoparticles at plasmonic resonance. $\mathbf{M}_{sca}$ and $\mathbf{M}_{abs}$ are the optical torque generated from scattering and absorption, respectively. **A.** The incident light can exert torque on a triangular particle through both scattering and absorption. **B.** In the case of a circular particle, scattering cannot contribute to an optical torque.

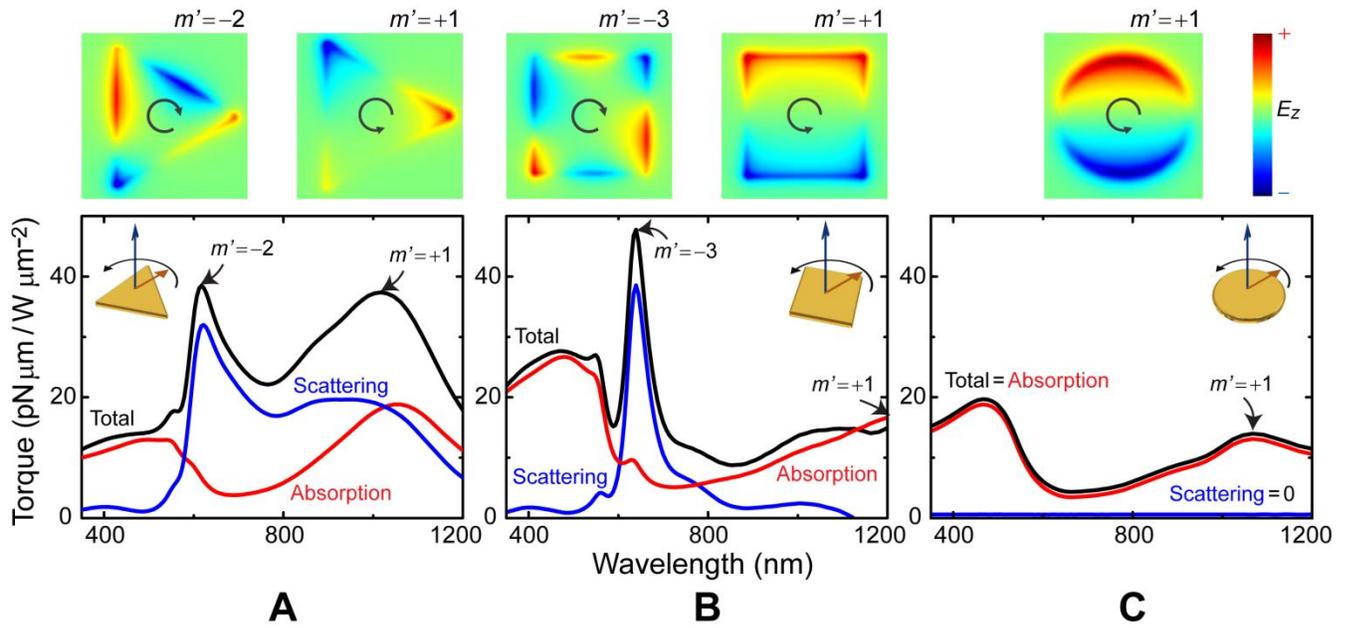

**Figure 2.** Calculated torque spectra for plasmonic particles excited at different orders of multipolar resonance. The total torque (black line) equals the summation of the scattering (blue) and absorption (red) contributions. Multipolar resonance peaks are labeled with the corresponding azimuthal order number $m'$. The corresponding electric field snapshot for each peak is plotted above, with black circular arrows denoting the direction in which the field pattern rotates. **A.** The triangular particle possessing 3-fold symmetry supports a dipole mode ($m'=1$) and a negative quadrupole mode ($m'=-2$). **B.** The square particle possessing 4-fold symmetry supports a dipole mode and a negative hexapole mode ($m'=-3$). **C.** In contrast, the circular particle only supports a dipole mode, and the no torque is generated from scattering. All three particles have the same lateral characteristic length of 400nm and thickness 40nm. Videos of the rotating electric fields are provided in the supplementary materials.

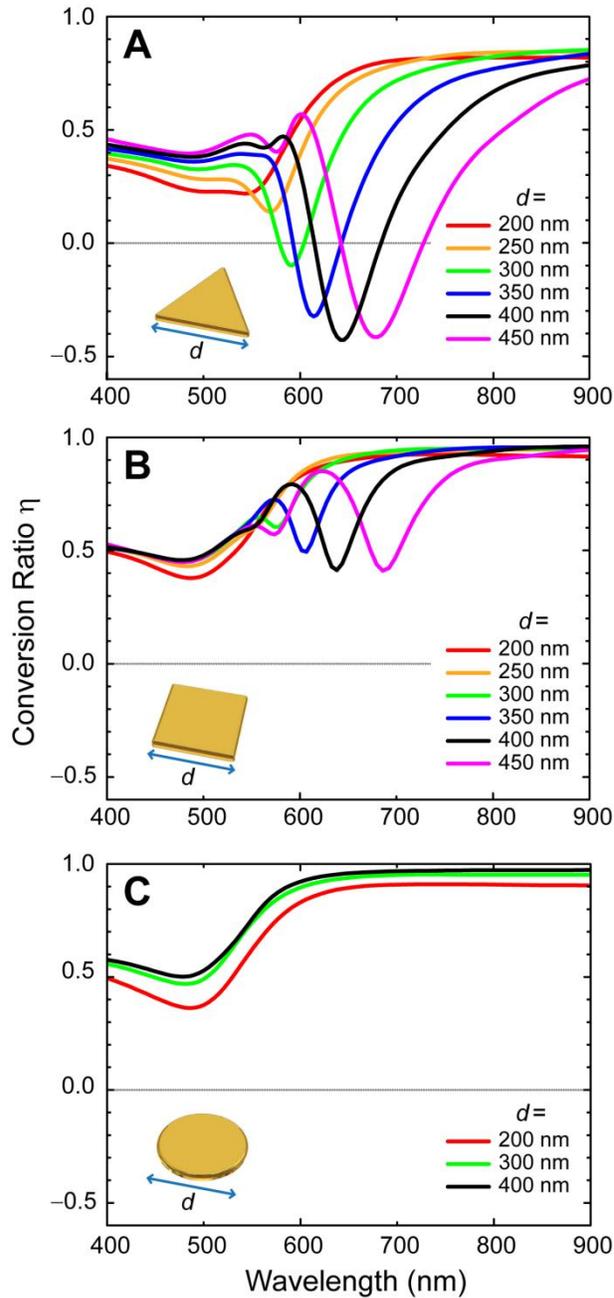

**Figure 3.** The angular momentum conversion ratio. The strength of angular momentum conversion is governed by the quality of multipolar plasmon resonance. The size parameter $d$ is varied in each spectrum around the optimum size. **A.** The triangular particle shows a dip in the conversion ratio at the negative quadrupole mode ($m' = -2$). **B.** The square particle shows a dip at the negative hexapole mode ($m' = -3$). **C.** The circular particle shows no multipolar conversion dip.

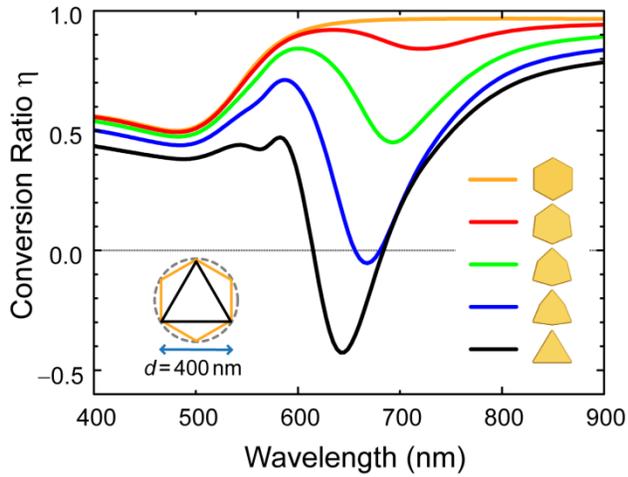

**Figure 4.** The importance of broken rotational symmetry in angular momentum conversion. The conversion ratio is plotted while a 6-fold hexagonal particle is gradually changed to a 3-fold triangular particle. $\eta$ of the hexagon (yellow line) is almost indistinguishable from that of a circular particle (refer to Figure 3). The efficiency of the negative conversion is enhanced by breaking the 6-fold symmetry, thereby creating a larger change in the optical angular momentum after scattering.

## ABBREVIATIONS

SAM spin angular momentum, OAM orbital angular momentum, SP surface plasmon, FDTD finite-difference time-domain, LCP left circular polarization, RCP right circular polarization.

## ACKNOWLEDGMENT

The authors acknowledge the financial support by NSF (CMMI Award No. 1120724) and AFOSR MURI (Award No. FA9550-12-1-0488). The authors thank Q. Hu, J. Xu, and A. Kumar for helpful discussions. YE Lee acknowledges Samsung Scholarship Foundation for additional financial support. KH Fung acknowledges the funding support from the Hong Kong RGC under the Early Career Scheme (Grant no. 509813).

## SUPPLEMENTARY MATERIALS

Permittivity and the material response (S1), optical extinction spectra (S2), a movie of the electric field distribution (S3).

# Supplementary information

**S1. Material Property**

The dielectric function of gold used in all our numerical simulations is plotted in Figure S5. The experimental data from Palik [1] has been smoothly fitted. The imaginary part of the dielectric function is connected to the absorption of the material.

The optical response of a nanostructure is governed by both the material response and the geometrical structure. It is important to distinguish between the two, in order to clearly understand the mechanism behind the resonantly enhanced transfer of angular optical momentum. According to the results in the main text, the enhanced optical torque from scattering is predominantly governed by the geometry of the nanoparticles, whereas that from absorption is nearly unaffected by the shape change. The absorptive behavior solely reflects the intrinsic response of the material. The same was found to be true for metallic particles made of other materials, such as silver and aluminum, for nanoparticles in the size range that supports multipolar plasmonic resonance.

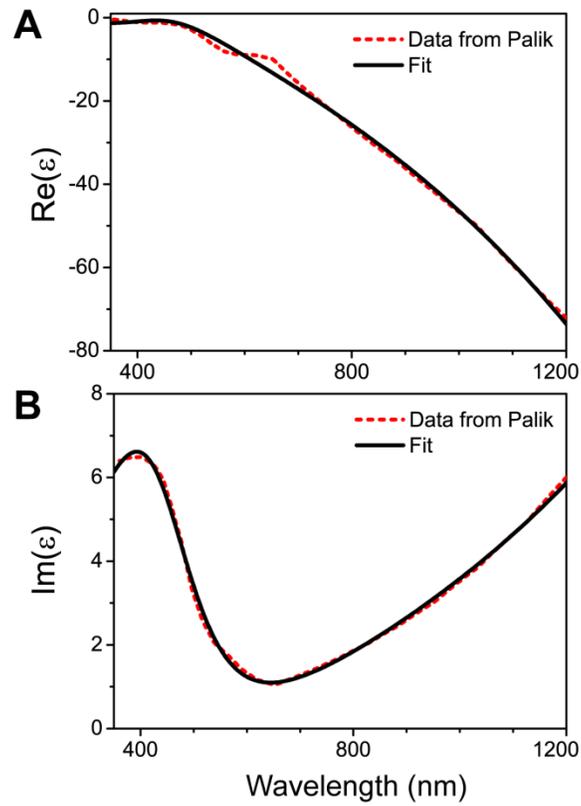

**Figure S5.** The dielectric function of gold. **A.** Real part of the dielectric function. **B.** Imaginary part of the dielectric function. The dotted line is the smoothed experimental data from Palik, and the solid like is the FDTD function fit.

## S2. Extinction Spectra

The extinction, scattering, and absorption spectra of the gold nanoparticles are plotted in Figure S6. . This represents the strength of the combined response from the material, size, and shape of the nanoparticles. The cross sections are calculated using the total-field scattered-field (TFSF) source in the FDTD simulation, which is a standard treatment. [2] The polarization of the incident light does not change the optical response because of the simple particle geometry. Without loss of generality, a monochromatic, normally incident, linearly polarized plane wave is used as the source of excitation.

The optical response for the gold nanoparticles with the characteristic length of $d = 400$nm is dominated by scattering, which means that the number of photons that are scattered is much greater than the number of photons absorbed. Nevertheless, the process of scattering has been longtime overlooked as a possible mechanism to create torque, since there are no means to significantly alter the angular momentum carried by the scattered field with conventional dielectrics and large particles.

The influence of the material response is prominent in the absorption cross section (red dotted curve), especially below the wavelength of 500nm. According to the three absorption cross sections, the amount of light absorbed is nearly the same for all three particles. While particles with the same material and size with varied shape show minimal difference in absorption, a variation in material or size creates large differences in absorption. [3] In other words, absorption is predominantly governed by the number of the charges on the metallic surface, rather than by a small change in their distribution.

It is worth highlighting the importance of the particle dimension by comparing the results from a 400nm particle and a 40nm particle. While both can exhibit plasmonic resonance, the nature of the localized surface plasmon resonance (LSPR) should be extremely different for the two cases. The optical response of the 40nm particle is almost entirely dominated by absorption, [4] unlike the 400nm particle in Figure S6. While surface plasmon resonance can be excited when there is a surface between a metal and a dielectric, the characteristics of this can be extremely different depending on the size of the particles. [5]

In contrast, the scattering cross section (blue curve) shows a clear difference in each spectrum in Figure S6, especially regarding the emergence of the small resonance peak near the wavelength of 500nm. The large peak above 900nm represents the dipole mode, and the small peak corresponds to the negative quadrupole mode for the triangle in Figure S6 (a), and to the negative hexapole mode for the square in Figure S6 (b). The circle in Figure S6 (c) does not support any higher-order multipolar resonance. The shape of the plasmonic particles determines the nature of scattering. [5] This is especially remarkable when the dimension of the particle is comparable to the wavelength of light, as the result of this letter indicates.

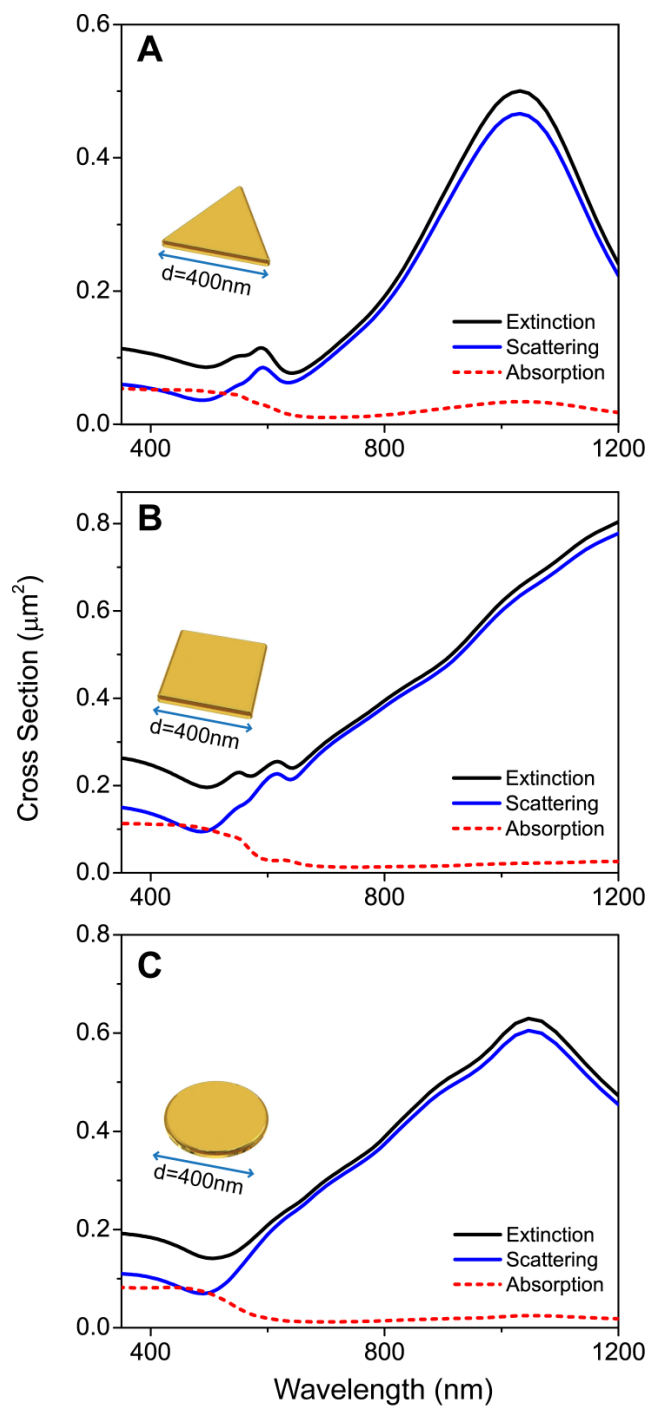

**Figure S6.** The extinction spectra. The extinction (black), scattering (blue), and absorption (red dotted) cross sections of the three characteristic geometries used in the main text are plotted as a function of wavelength. For all cases, the nanoparticles are made of gold; the characteristic lateral dimension is 400nm; and the thickness is 40nm. The dielectric environment is set as vacuum, with refractive index $n=1$.